\documentclass[aps,prd,nofootinbib,floatfix,superscriptaddress,
preprint
]{revtex4-1}
\usepackage{graphicx}% Include figure files
\usepackage{enumerate}% For non-numerical items
\usepackage{bm}% bold math
\usepackage{slashed}
\usepackage{xcolor}
\usepackage{braket}

\usepackage{amsmath}

\usepackage{amsfonts}

\usepackage{textcomp}

\DeclareMathOperator{\tr}{tr}

\begin{document}

\title{IR properties of chiral effects in pionic matter}

\author{A. Avdoshkin}
\email{avdoshkin@itep.ru}
\affiliation{ITEP, B. Cheremushkinskaya 25, Moscow, 117218, Russia}
\author{A. V. Sadofyev}
\email{sadofyev@lanl.gov}
\affiliation{ITEP, B. Cheremushkinskaya 25, Moscow, 117218, Russia}
\affiliation{Los Alamos National Laboratory, Theoretical Division, Los Alamos, NM 87545}
\author{V. I. Zakharov}
\email{vzakharov@itep.ru}
\affiliation{ITEP, B. Cheremushkinskaya 25, Moscow, 117218, Russia}
\affiliation{School of Biomedicine, Far Eastern Federal University, 690950, Vladivostok, Russia}

\begin{abstract}
Chiral effects exhibit peculiar universality in idealized theoretical limits. However, they are known to be infrared sensitive and get modified in more realistic settings. In this work we study how the corresponding conductivities vary with the constituent mass. We concentrate on a pionic realization of chiral effects which provides a better control over infrared properties of the theory. The pionic medium is considered at finite vector and axial isospin chemical potentials in the presence of an external magnetic field. This system supports electric and axial isospin currents along the magnetic field which correspond to chiral magnetic and chiral separation effects. We show that these currents are sensitive to the finite mass of the constituents but the conductivities follow a simple scaling with the corresponding charge densities as one would expect for polarization effects. It is argued that this relation can capture the dependence of chiral effects on other infrared parameters. Finally, we briefly comment on the realization of the 't Hooft matching condition in pionic media at finite densities.
\end{abstract}

\maketitle

\section{Introduction}
The axial anomaly is a one-loop-exact non-perturbative phenomenon closely tied with topological properties of the gauge field configuration. Inside a chiral medium anomaly may result in additional transports known as chiral effects which partially retain the anomalous universality: the corresponding conductivities have the same form in a wide variety of setups. For instance, chiral effects take the same form in a non-interacting fermion gas \cite{Vilenkin:1980fu, Vilenkin:1979ui} and in the opposite limit of hydrodynamics \cite{Son:2009tf}, for a recent review see \cite{Kharzeev:2015znc} and references therein. However, this is not the case away from the exact chiral limit, the anomalous transport is argued to have a strong dependence on the mass of the medium constituents and is infrared (IR) dependent in general, see e.g. \cite{Sadofyev:2010is, Hou:2011ze, Kirilin:2012mw, Kirilin:2013fqa, Gorbar:2013upa, Jensen:2013vta, Buividovich:2013hza, Yamamoto:2015fxa, Kaplan:2016drz}. Recent experimental progress in the search for the anomalous transport in condensed matter system \cite{Li:2014bha, Gooth:2017mbd} and the ongoing activity in heavy-ion experiments \cite{Kharzeev:2015znc} require further extension of the theoretical picture to more realistic setups incorporating various IR parameters and, particularly, the finite mass of constituents. 

Commonly chiral effects are discussed in the context of a chiral plasma made of massless Dirac fermions. In this medium  at finite vector and axial chemical potentials $\mu, \mu_5$ an external magnetic field results in electric and axial currents $J=\frac{\mu_5}{2\pi^2}B,~J_5=\frac{\mu}{2\pi^2}B$ known as chiral magnetic and chiral separation effects (CME and CSE correspondingly). In the strong field limit, one can count the contributions of different Landau levels to the transport \cite{Metlitski:2005pr} and the average axial current density is given by 
\begin{eqnarray}
J_5^i=\frac{1}{2\pi}n(T, \mu)B^i\,,
\label{eq:mass_dep_f}
\end{eqnarray}
where $n(T, \mu)$ is the fermion number density in the lowest Landau level (LLL). At zero temperature and non-zero $m$, the density has a simple analytic form resulting in
$$J_5^i=\frac{\sqrt{\mu^2-m^2}}{2\pi^2}B^i~~\xrightarrow[m\to 0]~~J_5=\frac{\mu}{2\pi^2}B\,$$
and in the massless limit one finds the standard expression for the CSE current. It should be mentioned that at $m\neq 0$ the diagrammatic relation of the CSE with the triangle anomaly is less direct. However, the CSE current is still saturated by LLL and is, in this sense, anomalous. 

There is a particularly interesting observation to be pointed out: according to (\ref{eq:mass_dep_f}) the corresponding conductivity follows a simple linear scaling with the fermion density as one would expect for a spin polarization \cite{Metlitski:2005pr}. Thus, the CSE mass dependence is fixed by the mass dependence of the density but it is unclear to what extent this relation could be generalized. For instance, in the case of a QCD medium it is known that both CME and CSE can take place not only in the high temperature limit of quark-gluon plasma, but also in the confined phase represented by a pionic medium at finite densities. The latter setup is considerably different from the one discussed above due to strong interactions which modifies IR properties of the theory. 

The anomalous transport in a pionic medium was studied previously, see e.g. \cite{Lublinsky:2009wr, Fukushima:2012fg, Kalaydzhyan:2014bfa, Huang:2017pqe}, and we extend those considerations to the presence of a finite pion mass. The pionic medium has a rich phase diagram \cite{Son:2000by, Aharony:2007uu, Son:2007ny} and provides a toy model to study chiral effects in distinct regimes. Such setup is additionally interesting due to recent proposals to search for polarization effects in hadronic products of heavy-ion collisions which can be generated in part by anomalous dynamics, see e.g. \cite{Rogachevsky:2010ys, Baznat:2013zx, Becattini:2013vja, Sorin:2016smp, Becattini:2016gvu} and references therein.  We show that in pionic media CME and CSE depend on the pion mass only through the corresponding charge densities similarly to the case of free fermions. In particular, we show that the mass dependence of CSE is captured by the mass dependence of the isospin density even in the condensed phase of isospin superfluidity. Finally, we comment on the realization of the 't Hooft matching condition in chiral media.

\section{Pion Condensation}
In this section we briefly review the behavior of the chiral theory of pions in the presence of finite densities but with no anomalous dynamics involved.  The corresponding low-energy effective field theory is well known and governed by the Chiral Lagrangian
\begin{eqnarray}
\mathcal L=\frac{1}{4}f_\pi^2\text{Tr}\left[\nabla_\mu U\nabla_\mu U^\dag-2\text{Re}~\text{Tr}~U\right]\,,
\label{chirallagrangian}
\end{eqnarray}
where $U\in SU(2)$ is the pion field and we introduce chemical potentials as zero components of vector and (auxiliary) axial gauge fields $\nabla_0 U=\partial_0 U-i\left([\mu_V, U]-\{\mu_A,U\}\right)$. 

To illustrate the effects of finite density, we concentrate on the case of non-zero isospin charge taking both chemical potentials in the $\tau_3$-direction, $\mu_{V(A)}=\frac{1}{2}\mu_{I,V(A)}\tau^3$. In the massless limit, the effective potential describing the vacuum state of the system takes the form
\begin{eqnarray}
V_{\text{eff}}=\frac{f_\pi^2}{8}\left[(\mu_{I,V}^2-\mu_{I,A}^2)\text{Tr}\left(\tau_3 U\tau_3 U^\dag\right)-2(\mu_{I,V}^2+\mu_{I,A}^2)\right]\,.
\label{veff}
\end{eqnarray}
The minimum of the potential (\ref{veff}) defines the vacuum alignment: if $\mu_{I,V}^2<\mu_{I,A}^2$ the vacuum solution lies in $(I, \tau_3)$-plane and $U_{vac}=e^{i\alpha}(I \cos\beta+i\tau^3\sin\beta)$, otherwise it is shifted to the $(\tau_1, \tau_2)$-plane and $U_{vac}=e^{i\alpha}(\tau_1 \cos\beta+i\tau_2\sin\beta)$. Isospin densities in the ground state are given by the variation of (\ref{veff}) with respect to the chemical potentials
\begin{eqnarray}
\rho_{I, V}=-\frac{f_\pi^2}{4}\left(\mu_{I, V}\text{Tr}\left(\tau^3 U\tau^3 U^\dag\right)-2\mu_{I, V}\right)\nonumber\\
\rho_{I, A}=-\frac{f_\pi^2}{4}\left(-\mu_{I, A}\text{Tr}\left(\tau^3 U\tau^3 U^\dag\right)-2\mu_{I, A}\right)
\label{densities}
\end{eqnarray}
and one can note that for $U$ in $(I, \tau_3)$-plane $\rho_{I, V}=0,~\rho_{I, A}=f_\pi^2\mu_{I,A}$ while for $U$ in $(\tau_1, \tau_2)$-plane $\rho_{I, V}=f_\pi^2\mu_{I,V},~\rho_{I, A}=0$, see e.g. \cite{Aharony:2007uu}.

At finite mass, the axial subgroup is explicitly broken and one cannot introduce the axial chemical potential since the charge is not conserved. Moreover, there is an additional scale in the problem and, if chemical potential $\mu_{I,V}$ is much smaller than the mass, one expects the system to stay in the trivial vacuum state $U=I$ while in the massless limit any $\mu_{I, V}$ results in a rotation of the vacuum. Indeed, for $|\mu_{I, V}|<m_\pi$ there are no particles in the vacuum, the isospin density is zero and the corresponding solution has to be trivial $U_{vac}=I$. On the other hand, in the limit $|\mu_{I, V}\gg m_\pi$ the system is expected to behave as in the massless case. This picture can be checked explicitly by studying the static limit of the Chiral Lagrangian for $m\neq0$. The ground state should minimize the potential energy
\begin{eqnarray}
V_{\text{eff}}=\frac{f_\pi^2}{8}\mu_{I,V}^2\text{Tr}\left[\tau^3 U\tau^3 U^\dag-1\right]-\frac{f_\pi^2 m_\pi^2}{2}\text{Re}~\text{Tr} U\,. \nonumber
\label{veffmass}
\end{eqnarray}
The pion filed minimizing (\ref{veffmass}) can be parametrized as $U=\cos\alpha+i(\tau_1\cos\phi+\tau_2\sin\phi)\sin\alpha$ at arbitrary $\mu_{I, V}$ reducing the previous expression to
\begin{eqnarray}
V_{\text{eff}}=\frac{f_\pi^2}{4}\mu_{I,V}^2(\cos 2\alpha-1)-f_\pi^2 m_\pi^2\cos\alpha\nonumber\,.
\end{eqnarray}
Following the discussion in \cite{Son:2000by}, one finds that the minimum of this potential with respect to $\alpha$ corresponds to $\cos\alpha=\frac{m_\pi^2}{\mu_{I,V}^2}$ which turns into $U_{vac}=I$ for $|\mu_{I,V}|<m_\pi$. This vacuum rotation corresponds to a condensation of $\pi^\pm$ for $|\mu_{I,V}|>m_\pi$ and the ground state can be seen as a pionic superfluid. Note that the new minimum of the potential is degenerate with respect to $\phi$ which corresponds to a Goldstone mode of the spontaneously broken isospin symmetry. The isospin density in the ground state is non-zero only if $|\mu_{I,V}|>m_\pi$, the expression (\ref{densities}) is unmodified by mass and one finds
\begin{eqnarray}
\rho_{I, V}=f_\pi^2\mu_{I, V}\left(1-\frac{m_\pi^4}{\mu_{I,V}^4}\right)\,,
\end{eqnarray}
which goes to zero for $|\mu_{I, V}|\to m_\pi$. This is a natural result since for $|\mu_{I,V}|<m_\pi$ there are no pions in the system at zero temperature. Note that naively one may consider $\mu_{I, A}$ to be an effective interaction ignoring the fact that the axial isospin charge is not conserved in the presence of the mass. In this case the system seemingly stays in the same vacuum even in the regime $|\mu_{I,V}|>m_\pi$. However, the non-conserved isospin axial charge is still well defined and one expects a state aligned with the $\tau_3$-direction due to finite density of $\pi^0$.

\section{Chiral Magnetic and Separation Effects}
Let us now consider anomalous transport in a pionic medium including the effect of the pion mass. The Chiral Lagrangian should be modified to include the axial anomaly which  is captured by the famous Wess-Zumino-Witten (WZW) term in this effective theory. In the case of $SU(2)$, the WZW term can be written in a particularly compact form \cite{Kaiser:2000gs, Kaiser:2000ck}
\begin{eqnarray}
\mathcal{L}_{WZW} = -\frac{N_c}{32\pi^2}\epsilon^{\mu\nu\rho\sigma} \bigg\{ \tr \bigg[U^{\dagger}\hat{r}_{\mu} U \hat{l}_{\nu} - \hat{r}_{\mu}  \hat{l}_{\nu}+i\Sigma_{\mu}(U^{\dagger}\hat{r}_{\nu}U+\hat{l}_{\nu})\bigg]\tr \left[ v_{\rho\sigma} \right] +\frac{2}{3}\tr \left[ \Sigma_{\mu}\Sigma_{\nu}\Sigma_{\rho}\right]\tr \left[ v_{\sigma} \right] \bigg\}\,,\notag\\
\label{WZW}
\end{eqnarray}
and the chemical potentials are included into the vector and axial fields. Here we use the standard notations
\begin{eqnarray}
&~&D_{\mu}U = \partial_{\mu}U -i r_{\mu}U +i Ul_{\mu}~~,~~\Sigma_{\alpha} = U^{\dagger}\partial_{\alpha}U,\nonumber\\ 
&~&r_{\mu} = v_{\mu}+a_{\mu}~~,~~l_{\mu} = v_{\mu}-a_{\mu},\label{chiraldefs}
\end{eqnarray}
and $\hat f=f-\frac{1}{2}\tr f$ is the traceless part of $f$. 

We are interested in currents sourced by (\ref{WZW}) which are expected to reproduce CSE and CME in the presence of the finite densities and an external magnetic field. Varying the action with respect to the axial field $a_{a, \mu}$, one can find the leading contribution to the axial current which is given by
\begin{equation}
J^{\mu}_{A, a}|_{WZW} =-\frac{N_c}{32 \pi^2}\epsilon^{\mu\nu\rho\sigma} \tr [ U^{\dagger} \tau_{a}U \hat{v}_{\nu}+U^{\dagger} \hat{v}_{\nu}U\tau_{a} - 2 \tau_{a} \hat{v}_{\nu} - i \Sigma_{\nu}(U^{\dagger}\tau_a U - \tau_a)]\tr v_{\rho\sigma}\,.
\label{JAWZW}
\end{equation}
%where we ignore contributions to $r_\mu,~l_\mu$ due to $B$ restricting our consideration to the weak field limit.
The background solution $U_{vac}$ is static since the chemical potentials $\mu_{V(A)}$ are explicitly taken into account in the definition of the covariant derivatives and one can omit the derivative term with  $\Sigma_\mu$, then
\begin{equation}
J^{i}_{A, a}|_{WZW} =-\frac{N_c\tr Q}{16\pi^2}\tr [U^{\dagger} \tau_{a}U \mu_V+U^{\dagger} \mu_V U\tau_{a} - 2 \tau_{a} \mu_V] B^i\,.
\end{equation}
This expression resembles the axial isospin density (\ref{densities}) and a direct substitution shows that the vacuum axial current is proportional to it
\begin{equation}
J^{i}_a|_{WZW}=\frac{N_c \tr Q}{4\pi^2 f_\pi^2}\rho_{I, V}B^i\,.
\label{CSE_pion_dens}
\end{equation} 
This is in analogy\footnote{The similarity in the structure of the CSE current and its IR dependence, however, cannot be used for a direct comparison of the responses in the two regimes at least within the current consideration.} with the strong field limit in the fermionic case (\ref{eq:mass_dep_f}) and the finite mass effects are incorporated similarly being accounted through the vacuum solution $U_{vac}$ which defines the ground state density. Note that the leading corrections to the background current due to perturbations around $U_{vac}$ are also captured by (\ref{CSE_pion_dens}) if the corresponding terms are taken into account in the density expression. 

An interesting illustration to this relation can be found in the phase of the pionic superfluidity which requires non-zero $|\mu_{I, V}|>m_\pi$. The condensation results in a non-trivial $U_{vac}$ which corresponds to non-zero isospin density in the ground state even at zero temperature. Then there is a vacuum axial current in the isospin component which reads
\begin{eqnarray}
J^{i}_{A, 3}|_{WZW}  = \frac{N_c\tr Q}{4 \pi^2}  \mu_{I, V}\left(1 -\frac{m_\pi^4}{\mu_{I, V}^4}\right)   B^{i}\,,
\end{eqnarray}
and the current disappears in the limit $(|\mu_{I, V}|-m_\pi)\to 0$ along with the density. We can see that the IR dependence of the CSE current is indeed captured by the corresponding charge density even in the presence of the condensate. 

In the phase of the pionic superfluidity, one has to be careful treating magnetic field effects since the ground state possesses some electric charge density unless it is removed by an external charge added to the system. Therefore it is instructive to consider the relation (\ref{CSE_pion_dens}) in a physically simpler setup involving no charge in the ground state. For $\mu^2 < m_\pi^2$ the vacuum solution is trivial, in this setup $J^{i}_{3}|_{WZW} =0$ at zero temperature following to (\ref{CSE_pion_dens}). However, as mentioned above, corrections to the isospin density at finite temperature should result in a non-zero isospin axial current. In the high temperature limit $ T \gg m_\pi$, one can ignore the mass of constituents and the leading contribution to $J^{i}_{A, 3}$ is given by
\begin{eqnarray}
J^{i}_{A, 3}|_{WZW} = \frac{N_c \mu_{I, V}}{4 \pi^2}  \frac{\pi_1^2 + \pi_2^2}{f_{\pi}^2} B^i =  \frac{N_c \mu_{I, V}}{24 \pi^2}  \frac{T^2}{f_{\pi}^2} B^i\,,
\end{eqnarray}
where we use the thermal expectation of the pion field $\left\langle \pi^2 \right\rangle = \frac{T^2}{12}$ (c.f. \cite{Lublinsky:2009wr, Kalaydzhyan:2014bfa}).

In the massless limit, the axial charge is classically conserved and one can derive the CME current within the same procedure introducing the axial chemical potential $\mu_{I, A}\neq 0$. We concentrate on the isospin component of the axial charge which is non-anomalous with respect to the strong interaction. The anomalous contribution to the electric current can be obtained by varying the WZW action (modified by the shift of the axial charge) with respect to the electric component of the vector field $v_\mu$, then
\begin{equation}
J^{i}_{V}|_{WZW} =\frac{N_c \tr Q}{16 \pi^2}\tr [U^{\dagger}\hat Q U \mu_A+U^{\dagger} \mu_A U\hat Q + 2\hat Q \mu_A] B^i\,,
\label{JVCME}
\end{equation}
where $\hat Q=\tau^3$ is the traceless part of the electric charge. Comparing (\ref{JVCME}) with (\ref{densities}), one finds that the CVE current is proportional to the axial isospin density $J^{i}_V|_{WZW}\sim \frac{1}{f_\pi^2}\tr\left[\rho_A\hat Q\right] B^i$. We substitute the isospin component of the axial chemical potential to (\ref{JVCME}) and take $U_{vac}=I$ then the CME current takes the form
\begin{gather}
J^{i}_{V}|_{WZW}  = \frac{N_c\tr Q}{4\pi^2}\mu_{I, A} B^i\
\end{gather}
which agrees with the previous considerations, see e.g. \cite{Fukushima:2012fg, Kalaydzhyan:2014bfa}. The finite mass effects explicitly violate conservation of all axial charge components and one has to be careful analyzing this regime. However, one expects that the relation $J^{i}_V|_{WZW}\sim\frac{1}{f_\pi^2}\tr\left[\rho_A\hat Q\right] B^i$ survives even in this regime at least as a leading contribution at small pion mass.

\section{'t Hooft matching condition}
It is  worth emphasizing that the infrared sensitivity of the matrix elements of the currents in the medium which we find is  in actually sharp contrast with more familiar examples from elementary particle physics. Consider first matrix element of the axial current over heavy states $|h_i\rangle, |h_f\rangle$. In the exact chiral limit:
\begin{equation}
\label{pole}
\langle h_f|j^5_{\alpha}|h_i\rangle~=~f_{\pi}\Big(\frac{q_{\alpha}q_{\beta}}{q^2}
-\delta_{\alpha\beta}\Big)
\langle h_f|\tilde{j}^5_{\alpha}|h_i\rangle\,,
\end{equation}
where $q_{\alpha}$ is
the 4-momentum carried in by the current 
$j^5_{\alpha}$ and
$\tilde{j}^5_{\alpha}$ is the axial current with the contribution of the Goldstone particle (pion) removed. The pole in the matrix element (\ref{pole}) is due to a virtual pion. Introduction of a finite pion mass $m_{\pi}$ results in the shift in the  position of the pole,
$$1/q^2~\to~1/(q^2+m_{\pi}^2)\,.$$
This shift of the pole is responsible for the whole of $m_{\pi}^2$ dependence in the infrared region.

Note that the matrix element (\ref{pole}) of the axial current contains  a pole and is infrared sensitive at small $q^2$. The matrix element of the divergence of the current $\partial^{\alpha}j_{\alpha}^5$ on the other hand reduces to a polynomial and is not infrared sensitive. Beginning with the seminal paper \cite{Son:2009tf}, the chiral effects are treated as higher order terms in the hydrodynamic expansion in derivatives. Thus, one could expect that the matrix elements of the currents are sensitive to specific hydrodynamic excitations \cite{Dolgov:1971ri}. Now we can check these expectations in the case of pionic superfluidity. We will see that the pattern of the infrared sensitivity is very different in fact from the elementary-particle case.

For the pion medium, there are real pions present and, as we have mentioned a few times, axial current itself is proportional to the density of pions $\rho_{\pi}$. In the limit of exact symmetry, the matrix element of the axial current over the pion medium is given by:
\begin{equation}\label{medium}
\langle medium| j_{\alpha}^5|medium\rangle~\approx~f_{\pi}\partial_{\alpha}\pi~=~
\rho_{\pi}u_{\alpha}\,,
\end{equation}
Obviously, in this case there is no pole due to the pion in the matrix element of the axial current.  What is changed with introduction of a finite pion mass is the expression for the pion density $\rho_{\pi}$.

Proceeding to loop effects, the matrix element associated with the anomalous triangle graph perturbatively is given by:
\begin{equation}\label{thooft1}
\langle0|j_{\alpha}^5|\gamma\gamma\rangle_{pert}~=~\frac{q_{\alpha}}{q^2}\frac{\alpha_{el}}{2\pi}(Q_u^2-Q_d^2)N_cF_{\alpha\beta}\tilde{F}^{\alpha\beta}\,,
\end{equation}
where $Q_u,Q_d$ are charges of the light quarks, $N_c$ is the number of colors. Note that higher orders in perturbation theory vanish due to the Adler-Bardeen theorem \cite{Adler:1969er}. Phenomenologically, at small $q^2$ the same matrix element is given by
\begin{equation}\label{thooft2}
\langle0|j_{\alpha}^5|\gamma\gamma\rangle_{phen}~=~\frac{q_{\alpha}}{q^2}f_{\pi}f_{\pi^0\to \gamma\gamma}F_{\alpha\beta}\tilde{F}^{\alpha\beta}\,, 
\end{equation}
where $f_{\pi\gamma\gamma}$ is the constant of the $\pi^0~\to~\gamma\gamma$ decay, $F\tilde{F}~=~(1/2)\epsilon_{\alpha\beta\gamma\delta}F^{\alpha\beta}F^{\gamma\delta}$, and $F^{\alpha\beta}$ is the electromagnetic field strength tensor. Equating (\ref{thooft1}) and (\ref{thooft2}) reproduces the 't Hooft matching condition \cite{tHooft:1979rat} which fixes the product $f_{\pi}f_{\pi\gamma\gamma}$ in terms of $\alpha_{el}$.

Now we have a field theoretic language for the interaction of pions with heavy states and are, therefore, in position to evaluate loop corrections to (\ref{medium}). To this end we couple heavy states entering the matrix element (\ref{pole}) to the electromagnetic fields through loop graphs. These loop graphs are convergent at distances of order of inverse nucleon mass (as far as the nucleon is a ``typical '' heavy state). Moreover, such graphs have been extensively studied in connection with pion decay into two photons. All such graphs produce an effective action:
\begin{equation}\label{action}
\delta S~=~f_{\pi^0\to \gamma\gamma}\pi^0 F_{\alpha\beta}\tilde{F}^{\alpha\beta}~.
\end{equation}
Varying this effective action with respect to the electromagnetic potential $A_{\alpha}$ one finds a contribution to the electromagnetic current (known as the Wilzcek-Goldstone current: \cite{Goldstone:1981kk}),
\begin{equation}\label{polynomial}
j_{\alpha}^{el}~=~4f_{\pi^0\to \gamma\gamma}
\epsilon_{\alpha\beta\gamma\delta}\partial^{\beta}\pi^0
\partial^{\gamma}A^{\delta}
\end{equation} 
and, upon substitution $\partial^0\pi^0=f_{\pi}\mu^5$, we reproduce the current responsible for the chiral magnetic effect. The reason is that the interaction of pions with electromagnetic fields (\ref{action}) satisfies constraints due to the anomaly, or the the original 't Hooft matching condition, see above. 

It is worth emphasizing that in the case of pionic superfluidity the chiral magnetic effect does not reduce to higher orders in the hydrodynamic expansion. Indeed, the basic quantity determining the hydrodynamic expansion is the free-path length which is limited by collisions of constituents of the fluid. The current (\ref{polynomial}), on the other hand, is determined by the dynamics of heavy, virtual particles which are not constituents of the fluid, see also \cite{Teryaev:2017nro}. Thus, the chiral magnetic effect is associated in the considered model with the short distances, $r\sim m_N^{-1}$, and corresponds to a polynomial in the effective action (\ref{action}). It is not sensitive to the infrared physics at smaller scales.

Therefore, in the case of the hydrodynamic chiral magnetic effect the 't Hooft matching condition reduces in fact to its original form. In the case of the chiral vortical effect, the divergence of the corresponding piece in the axial current does not vanish on pure algebraic grounds but is rather proportional to
\begin{equation}\label{acceleration}
\partial^{\alpha}(\mu^2\epsilon_{\alpha\beta\gamma\delta}u^{\beta}\partial^{\gamma}
u^{\delta})~\sim~\mu^2(\vec{\Omega}\cdot \vec{a})\,,
\end{equation}
where $\vec{\Omega}$ is the angular velocity and $\vec{a}$ is acceleration of the fluid \cite{Sadofyev:2010is}. However, as is well known, introduction of the chemical potential $\mu$ cannot modify the divergence of the anomalous axial current. Therefore, the extra term (\ref{acceleration}) is to vanish because of the equation of motion of the fluid, for further details see \cite{Zakharov:2016lhp}. This constraint can be considered as a specific realization of the 't Hooft matching condition in hydrodynamics.

\section{Discussions}
Chiral effects are known to be IR sensitive and, particularly, one has to carefully study their behavior away from the exact massless limit. While the anomalous transport is commonly discussed in a weakly interacting fermionic system corresponding to the deconfined phase of QCD, it is also known that chiral effects can be generalized to a pionic medium. The two setups are considerably different in the IR limit but due to the universality of the anomaly one expects the anomalous transport to be similar. We compare these realizations of chiral effects to gain further insights into their IR sensitivity.

In this paper, we focus on the mass dependence of the CME and CSE currents in a pionic medium extending the previous considerations \cite{Lublinsky:2009wr, Fukushima:2012fg, Kalaydzhyan:2014bfa} beyond the exact chiral limit. It is known that in the deconfined phase the CSE current depends on the fermion mass only through the a certain fermion number density  (\ref{eq:mass_dep_f}). We show that it is also the case for the pionic realization of the anomalous transport. The CME ans CSE conductivities depend on the pion mass only through the corresponding densities and the currents simply scale with them in full analogy with the anomalous transport in the deconfined phase. In particular, we show that the mass dependence of CSE in the isospin axial current is captured by the mass dependence of the vector isospin density even in the condensed phase of isospin superfluidity \cite{Son:2000by, Aharony:2007uu, Son:2007ny}. In the massless limit, our results agree with the previous considerations of the anomalous transport in pionic media \cite{Lublinsky:2009wr, Fukushima:2012fg, Kalaydzhyan:2014bfa}. 

The replacement described above of the quark mass by the pion mass in CME and CSE conductivities is well expected. Indeed, the two masses are related and, moreover, pions are the only low energy degrees of freedom present in the system below the phase transition. On the other hand, the conductivities could, in principle, gain radiative corrections due to the strong interaction and in this sense the same form of the relation between the currents and the corresponding densities in deconfined and confined phases is a non-trivial result. One should also note that the transition to the pion mass makes the chiral effect less IR sensitive. Indeed, the anomalous transport is shown to be factorized to a density and an anomalous process. The triangle diagram for quarks in the deconfined phase depends on the photon virtuality of order $m_q^2$ while the dependence of the $\pi\to\gamma\gamma$ decay appears at order $k^2/m_\rho^2$. Thus, the IR sensitivity of the anomalous transport is considerably weaker within the pionic realization corresponding to its polynomial nature.

Finally, we comment on the realization of the 't Hooft matching condition in chiral media. The vortical contribution to the divergence of the axial current seemingly modifies the axial anomaly. We argue that the `t Hooft matching condition (\ref{thooft2}) has to stay unmodified in the hydrodynamic setup and one has to require the vortical contributions in the divergence to be zero (\ref{acceleration}). We stress that it is the case in the ideal hydrodynamics which is the proper description of a pionic medium in the hydrodynamic limit.

\section{Acknowledgments}
Authors are grateful to O. Teryaev, V. Kirilin, and S. Sen for interesting discussions. The work on sections III and IV is supported by Russian Science Foundation Grant No. 16-12-10059. The work of AS is partially supported through the LANL/LDRD Program.

\bibliographystyle{unsrt}
\bibliography{pions.bib}

\end{document}